\documentclass{article}
\usepackage{spconf,amsmath,graphicx,amssymb, times, multirow}

\usepackage{algorithm}
\usepackage{algpseudocode}
\makeatletter
\def\BState{\State\hskip-\ALG@thistlm}
\makeatother
\algdef{SE}[SUBALG]{Indent}{EndIndent}{}{\algorithmicend\ }%
\algtext*{Indent}
\algtext*{EndIndent}

\usepackage{amssymb}

\usepackage{graphicx,caption}

\usepackage{array}
\usepackage{lipsum}
\usepackage{float}
\usepackage{rotating}

\begin{document}
\title{ Logarithmic Frequency Scaling and Consistent Frequency Coverage for the Selection of Auditory Filterbank Center Frequencies }
\name{Shoufeng Lin}
\address{Department
of Electrical and Computer Engineering, Curtin University \\ Kent Street, Bentley, Perth, Western Australia, 6102 \\
shoufeng.lin@postgrad.curtin.edu.au; ee.linsf@gmail.com }

\maketitle

\begin{abstract}
This paper provides new insights into the problem of selecting filter center frequencies for the auditory filterbanks. 
We propose to use a constant frequency distance and a consistent frequency coverage as the two metrics that motivate the logarithmic frequency scaling and a regularized selection of center frequencies. The frequency scaling and the consistent frequency coverage have been derived based on a common harmonic speaker signal model. Furthermore, we have found that the existing linear equivalent rectangular bandwidth (ERB) function as well as any possible linear ERB approximation can also lead to a consistent frequency coverage. The results are verified and demonstrated using the gammatone filterbank. 
\end{abstract}

\begin{keywords}
auditory filterbank, speech signal processing, frequency scaling, frequency coverage, ERB.
\end{keywords}

\section{Introduction}
\label{intro}
%

{A}{uditory} filterbanks have been widely accepted and applied in numerous speech signal processing algorithms especially in the computational auditory scene analysis (CASA) area \cite{CASAwang}, for various applications including the speech enhancement, recognition and transcription. 

A typical auditory filterbank consists of two parts, i.e. the filter type and the centre frequencies of filters. Common filter types include the gammatone, gammachirp, and their variants \cite{holdsworth1988implementing}, which simulate the auditory response of human hearers. Choice of center frequencies of the auditory filters has evolved from the earlier critical bandwidth and the critical-band-rate scale \cite{zwicker1980analytical}, to the polynomial approximation of equivalent rectangular bandwidth (ERB) \cite{moore1983suggested}, and the currently well-accepted linear ERB \cite{glasberg1990derivation}, as well as their corresponding ERB-rate scales (ERBS). 

Although the linear ERB approximation in \cite{glasberg1990derivation} has been found useful in practical implementations, it has been based on experimental findings through psychoacoustic measurement and curve-fitting. 
Logarithmic frequency scales have also been applied \cite{sun2002pitch, nolan2003intonational, biesmans2017auditory}.  However, the selection of the number of subbands for a given frequency range still remains empirical for both of the ERB rate scale and the logarithmic scale. 

In this paper, we further investigate the frequency scaling and provide new insights including a new proposed frequency coverage metric, and also derivations of a new frequency scaling function that lead to consistent frequency coverage for auditory filterbanks. Moreover, based on the proposed definition of frequency coverage, we also derive an expression for the frequency coverage metric from the existing linear ERB. 

\section{Equivalent Rectangular Bandwidth Scale}
\label{section:erbs}
The ERB of a particular filter is defined as the bandwidth of a rectangular filter to pass the same energy of the filter \cite{moore1983suggested, glasberg1990derivation}. The relationship between the ERB of the human auditory filter and the center frequency has been studied extensively using analytical expressions to approximate measurement data from psychoacoustic experiment. 
An early approximation has the polynomial form \cite{moore1983suggested}
\begin{equation}
\widehat{\mathrm{ERB}}(f) = a \cdot f^2 + b\cdot f + c ,
\end{equation}
where $f$ is the frequency in unit of Hz, and $a, b, c \in \mathbb{R}$ are parameters. 
However, one of the most widely accepted analytical approximation over the past decades has been the linear form \cite{glasberg1990derivation}
\begin{equation} \label{eq:ERB}
\begin{aligned}
\widetilde{\mathrm{ERB}}(f) = & 24.7 \cdot (0.00437\cdot f + 1) \\
= & 24.7 + 0.108 \cdot f .  
\end{aligned}
\end{equation}
Each ERB corresponds to a constant distance along the basilar membrane \cite{moore1986parallels, glasberg1990derivation} in cochlea. 
%

The ERB-rate scale (ERBS) has been developed to scale frequency in terms of units of the ERB, by solving the integral \cite{moore1983suggested, glasberg1990derivation}:
\begin{equation} \label{eq:ERBSfunc}
\widetilde{\mathrm{ERBS}}(f) = \int \frac{1}{\widetilde{\mathrm{ERB}}(f)} df ,
\end{equation}
with the boundary condition
\begin{equation} \label{eq:ERBSbdr}
\widetilde{\mathrm{ERBS}}(0) = 0 . \\
\end{equation}
Using (\ref{eq:ERB}) in (\ref{eq:ERBSfunc}) and (\ref{eq:ERBSbdr}) yields \cite{glasberg1990derivation} 
\begin{equation} \label{eq:ERBSfun}
\widetilde{\mathrm{ERBS}}(f) = 21.4 \cdot \lg (0.00437\cdot f + 1) .
\end{equation}

The ERB and ERBS given in (\ref{eq:ERB}) and (\ref{eq:ERBSfun}) have been applied in numerous auditory studies, for selecting the center frequencies of the auditory filterbank \cite{patterson1987efficient}, yet the ERB approximation is still found as a result of curve-fitting from experiments, and the number of subbands for a given frequency range is still an empirical parameter. 

\section{Suggested Frequency Scaling and Coverage}

\subsection{Speaker Signal Model}
Based on the source excitation - vocal tract models for the process of speech production \cite{deller1993discrete}, as well as the amplitude-modulation (AM) and frequency modulation (FM) structure \cite{maragos1993energy}, a harmonic model is used for the speaker signal:
%
\begin{equation}
\label{eq:speechModel}
s_q(t) = \sum \limits _{{\hbar}=1} ^{H_q} s^{(\hbar)}_q(t) , 
\end{equation}
\begin{equation}\label{eq:speechModel2}
s^{(\hbar)}_q(t) = A^{(\hbar)}_{q}(t) \cdot \cos \big( {\hbar} \cdot \omega_q \cdot t + \phi^{({\hbar})}_{q}(t) \big) ,
\end{equation}
where $t \in \mathbb{R}$ is continuous time, $s_q(t)$ the speech signal from the $q$-th speaker, $q = 1,...,Q $, integer $Q\geq 1$ the number of concurrent speakers,  $s^{(\hbar)}_q(t)$ the $\hbar$-th harmonic of speaker $q$, integer ${\hbar}$ the order of harmonics for a speaker, integer $H_q$ the maximum order of harmonics for speaker $q$, $A^{({\hbar})}_{q}(t) \geq 0$ the envelope of each harmonic, $\phi^{({\hbar})}_{q}(t) \in \mathbb{R}$ the phase (which is short-time constant for speech signals), and $\omega_q > 0$ the (angular) fundamental frequency. 

With appropriate selection of filter center frequencies, the auditory filterbank ideally separates into subbands the harmonic components of not only a single speaker, but also multiple concurrent speakers, based on the time-frequency sparsity assumption of speech signals \cite{yilmaz2004blind}.

\subsection{Logarithmic Frequency Scaling}

In practice, concurrent speakers usually have different fundamental frequencies. Thus we can denote fundamental frequencies of two speakers as $f_1$, $f_2$ ($f_1 = \omega_1/2\pi$, $f_2 = \omega_2/2\pi$, $f_1 \neq f_2$), and their difference is
\begin{equation}
\Delta f = f_1 - f_2 .
\end{equation} 
Thus from (\ref{eq:speechModel2}) the frequency difference of their ${{\hbar}}$-th harmonic is ${{\hbar}} \cdot \Delta f$. This means that their harmonics (of same order) are more distant at higher frequencies on the linear frequency scale, which makes selection of the filterbank center frequencies difficult for a regular per-speaker estimate. 

We thus propose a frequency scaling function $\Upsilon(\cdot)$ that satisfies (\ref{eq:freqscale}) so that speech components of separate speakers appear equidistantly, with respect to (w.r.t.) $\hbar$ : 
\begin{equation}
\label{eq:freqscale}
\Upsilon({{\hbar}} \cdot f_1) - \Upsilon({{\hbar}} \cdot f_2) \equiv \mathit{Constant},\,\mathrm{w.r.t.}\,{{\hbar}} .
\end{equation}
%

The logarithmic functions are functional solutions to (\ref{eq:freqscale}):
\begin{equation} \label{eq:logfunc}
{\Upsilon}(\cdot) = A\cdot\mathrm{log}_B(\cdot) +C  ,
\end{equation} 
where $A>0,~B>1,~C \in \mathbb{R}$. They also have better resolutions for the lower frequencies, which aligns with the fact that most speech energy falls in low frequencies (e.g. fundamental frequencies and their lower-order harmonics). We can easily verify from (\ref{eq:logfunc}) that ${\Upsilon}({{\hbar}} \cdot f_1) - {\Upsilon}({{\hbar}} \cdot f_2) \equiv A\cdot \big(\mathrm{log}_B (f_1 / f_2)  \big) $, which is constant with respect to $\hbar$.

Denote the ratio of center frequency to the bandwidth as $\eta_B^{(b)}$ for filter band $b$ ($b=1,...,N_b$), integer $N_b > 1$ is the number of filter bands, i.e. 
\begin{equation} \label{eq:Qfactor}
f_C^{(b)} = \eta_B^{(b)} \cdot f_B^{(b)} ,
\end{equation}
where $f_B^{(b)}$ and $f_C^{(b)}$ denote the bandwidth and center frequency of filter band $b$, respectively. 
$\eta_B^{(b)}$ is also referred to as the quality factor (Q-factor) of subband $b$. 

Denote the frequency range that we are interested in as $[f_{min}, f_{max}]$, where $f_{max} > f_{min} >0$. Assuming that the center frequencies of filter bands are equidistantly spaced in the proposed frequency range, we have
\begin{equation} \label{eq:upsilon0}
\Upsilon(f_C^{(b)}) = \frac{(N_b-b) \cdot \Upsilon(f_{min}) + (b-1) \cdot \Upsilon(f_{max})}{N_b-1} ,
\end{equation}
and
\begin{equation} \label{eq:centerfreqperband}
f_C^{(b)} = \Upsilon^{-1} \big( \Upsilon(f_C^{(b)}) \big) ,
\end{equation}
where ${\Upsilon}^{-1}(\cdot)$ denotes the inverse function of ${\Upsilon}(\cdot)$.

From (\ref{eq:logfunc}), (\ref{eq:upsilon0}) and (\ref{eq:centerfreqperband}), we can get for the new logarithmic frequency scaling 
\begin{equation} \label{eq:fcb}
\begin{aligned}
f_C^{(b)} & = \Upsilon^{-1} (\Upsilon(f_C^{(b)}) \\
& = \Upsilon^{-1} ( \frac{(N_b-b) \cdot \Upsilon(f_{min}) + (b-1) \cdot \Upsilon(f_{max})}{N_b-1}  ) \\
& = B ^ { \frac{ \frac{(N_b-b) \cdot \Upsilon(f_{min}) + (b-1) \cdot \Upsilon(f_{max})}{N_b-1}  -C}{A}} \\
\end{aligned}
\end{equation}

\subsection{Proposed Frequency Coverage} \label{section:freqcoverage}
The auditory filterbank requires sufficient frequency coverage to capture all harmonic components of concurrent speakers. Here we propose to define the frequency coverage of the filterbank on the proposed frequency scale as
\begin{equation} \label{eq:freqcoverage}
\eta_C^{(b)} \triangleq \frac{\bar{\Sigma} f_{B} ^{(b)}}{\Delta f_C^{(b)}} ,
\end{equation}
where $\Delta f_C^{(b)}$ and $\bar{\Sigma} f_{B} ^{(b)}$ denote the distance between consecutive filter bands and the half of the sum of their bandwidths, as shown in (\ref{eq:deltafc}) and (\ref{eq:sumfb}), respectively:
\begin{equation} \label{eq:deltafc}
\Delta f_C^{(b)} \triangleq f_C^{(b+1)} - f_C^{(b)} ,
\end{equation}
and
\begin{equation} \label{eq:sumfb}
\bar{\Sigma} f_{B} ^{(b)} \triangleq \frac{1}{2} \cdot ( f_B^{(b+1)} + f_B^{(b)} ) .
\end{equation}
Apparently $\eta_C^{(b)} = 1$ gives a full coverage for ideal ``brick-wall'' bandpass filters with no overlap. For a practical auditory filterbank however, the filters always have finite roll-off rate, thus reasonable overlap is required for full coverage, leading to $\eta_C^{(b)} \approx 1$. Also depending on applications, we may have $\eta_C^{(b)} < 1$ when full coverage is not required.

Therefore from (\ref{eq:Qfactor}), (\ref{eq:fcb}) and (\ref{eq:freqcoverage}), we have when $\eta_B^{(b)} = \eta_B^{(b+1)}$, 
%
%
\begin{equation} \label{eq:etaCb}
\begin{aligned}
\eta_C^{(b)} &
= \frac{1}{2} \cdot \frac{ f_B^{(b+1)} + f_B^{(b)} }{ f_C^{(b+1)} - f_C^{(b)} } \\
& = \frac{1}{2 \eta_B^{(b)}} \cdot \frac{ f_C^{(b+1)} + f_C^{(b)} }{ f_C^{(b+1)} - f_C^{(b)} } \\
& = \frac{1}{2 \eta_B^{(b)}} \cdot \frac{ f_C^{(b+1)} / f_C^{(b)} + 1 }{ f_C^{(b+1)} / f_C^{(b)} - 1 }  \\
& = \frac{1}{2 \eta_B^{(b)}} \cdot \frac{ B^{ \frac{\Upsilon(f_{max}) - \Upsilon(f_{min})} {A (N_b-1)}} + 1 }{ B^{ \frac{\Upsilon(f_{max}) - \Upsilon(f_{min})} {A (N_b-1)}} - 1 } \\
& = \frac{1}{2 \eta_B^{(b)}} \cdot \frac{(\frac{f_{max}}{f_{min}})^{\frac{1}{N_b-1}}+1}{(\frac{f_{max}}{f_{min}})^{\frac{1}{N_b-1}}-1} ,
\end{aligned}
\end{equation}
%
which clearly shows that the frequency coverage on the logarithmic frequency scaling is consistent over the frequency range, i.e. if the Q-factor $\eta_B^{(b)}$ is a constant w.r.t subband index $b$, the resulting $\eta_C^{(b)}$ is also a constant value.

%
%
%
%


\subsection{Frequency Coverage of the Existing ERBS}

The existing ERB function (\ref{eq:ERB}) does not lead to a constant $\eta_B^{(b)}$, here we investigate its corresponding frequency coverage by applying the definition in (\ref{eq:freqcoverage}). 

Denote the general form of ERB in (\ref{eq:ERB}) as
\begin{equation} \label{eq:linearERB}
\hat{\upsilon} (f) = D + E \cdot f , 
\end{equation}
where $D \geq 0$, $E >0$. When $D = 24.7$, and $E = 0.108$ we have (\ref{eq:ERB}). 

The resulting ERBS following the process of (\ref{eq:ERBSfunc}) and (\ref{eq:ERBSbdr}) becomes:
\begin{equation}
\hat{\Upsilon}(f) = E' \lg ( 1+ D' \cdot f ) ,
\end{equation}
where 
\begin{equation}
D' \triangleq \frac{E}{D} ,
\end{equation}
and
\begin{equation}
E' \triangleq \frac{1}{E \cdot \lg e } . 
\end{equation}

Assuming the filter bandwidth is a constant scale of the ERB, which is true for some auditory filters, e.g. the gammatone filter \cite{holdsworth1988implementing}, i.e.
\begin{equation} \label{eq:constantK}
f_B^{(b)} = K \cdot \hat{\upsilon} (f) , 
\end{equation}
where $K > 0$ is a constant. Note here that the Q-factor is not constant as $D \neq 0$. 

Therefore, selecting equidistantly on the scale $\hat{\Upsilon}(f)$, similar to (\ref{eq:fcb}), we have 
\begin{equation} \label{eq:fcERBS}
\begin{aligned}
& f_C^{(b)} = \hat{\Upsilon}^{-1} (\hat{\Upsilon}(f_C^{(b)}) \\
& = \hat{\Upsilon}^{-1} ( \frac{(N_b-b) \cdot \hat{\Upsilon}(f_{min}) + (b-1) \cdot \hat{\Upsilon}(f_{max})}{N_b-1}  ) \\
& = \frac{1}{D'} \Big[ (1+D' f_{min})^{(N_b-b)} \cdot (1+D' f_{max})^{(b-1)} \Big] ^{\frac{1}{N_b-1}} \!\!\! - \! \frac{1}{D'} .
\end{aligned}
\end{equation}
Thus from (\ref{eq:freqcoverage}) and (\ref{eq:linearERB}) we have 
\begin{equation} \label{eq:etaCpsi}
\begin{aligned}
& \eta_{C,\hat{\upsilon}}^{(b)} = \frac{1}{2} \cdot \frac{ f_B^{(b+1)} + f_B^{(b)} }{ f_C^{(b+1)} - f_C^{(b)} } \\
& = K \cdot \frac{ D + \frac{E}{2} \cdot ( f_C^{(b+1)} + f_C^{(b)} ) }{ f_C^{(b+1)} - f_C^{(b)} } \\
& = \frac{E \!\! \cdot \!\! K}{2} \cdot \Big[ \Big( (1+D' f_{min})^{(N_b-b-1)} \!\! \cdot \!\! (1+D' f_{max})^{(b)} \Big) ^{\frac{1}{N_b-1}} \!\! + \\
& ~~~~~~~ \Big( (1+D' f_{min})^{(N_b-b)} \cdot (1+D' f_{max})^{(b-1)} \Big) ^{\frac{1}{N_b-1}}   \Big] \Big/ \\ 
&~~~~~~~~ \Big[ \Big( (1+D' f_{min})^{(N_b-b-1)} \cdot (1+D' f_{max})^{(b)} \Big) ^{\frac{1}{N_b-1}} - \\
&~~~~~~~~~ \Big( (1+D' f_{min})^{(N_b-b)} \cdot (1+D' f_{max})^{(b-1)} \Big) ^{\frac{1}{N_b-1}}   \Big] \\
& = \frac{E \cdot K}{2} \cdot \Big[ (1+D' f_{max})^{\frac{1}{N_b-1}}   +  (1+D' f_{min})^{\frac{1}{N_b-1}}   \Big] \Big/ \\ 
&~~~~~~ \Big[ (1+D' f_{max})^{\frac{1}{N_b-1}}  - (1+D' f_{min})^{\frac{1}{N_b-1}}   \Big] \\
& = \frac{E \cdot K}{2} \cdot \frac{(\frac{D + E\cdot f_{max}}{D + E \cdot f_{min}})^{\frac{1}{N_b-1}}+1}{(\frac{D + E \cdot f_{max}}{D + E \cdot f_{min}})^{\frac{1}{N_b-1}}-1} ,
\end{aligned}
\end{equation}
which is also constant over filter subbands. 
Thus as long as the ERB has the linear form as (\ref{eq:linearERB}) and assuming that (\ref{eq:constantK}) holds, the resulting frequency coverage is constant over frequency at given $f_{min}$, $f_{max}$ and $N_b$. 
Thus the number of subbands for a given frequency range $N_b$ can be derived from the required frequency coverage using (\ref{eq:etaCpsi}), and the subband center frequencies can then be calculated from (\ref{eq:fcb}) or (\ref{eq:fcERBS}). 

\section{Numerical Studies}

\subsection{New ERB and ERBS Functions}

From (\ref{eq:logfunc}) we have a new frequency scaling function that can lead to consistent frequency coverage for the auditory filterbank, as well as a constant Q-factor. 
Now we calculate the parameters.

Denote the maximum inaudible frequency as $f_{m}$, usually $f_{m} \approx 20$Hz, we use the boundary condition 
\begin{equation}
{\Upsilon}(f_{m}) = 0 , 
\end{equation}
instead of (\ref{eq:ERBSbdr}). Thus from (\ref{eq:logfunc}) we have
\begin{equation}
C = - A \cdot \log _B (f_m) . 
\end{equation}

\begin{figure}[!h]
\centering
\includegraphics[width=0.5\textwidth]{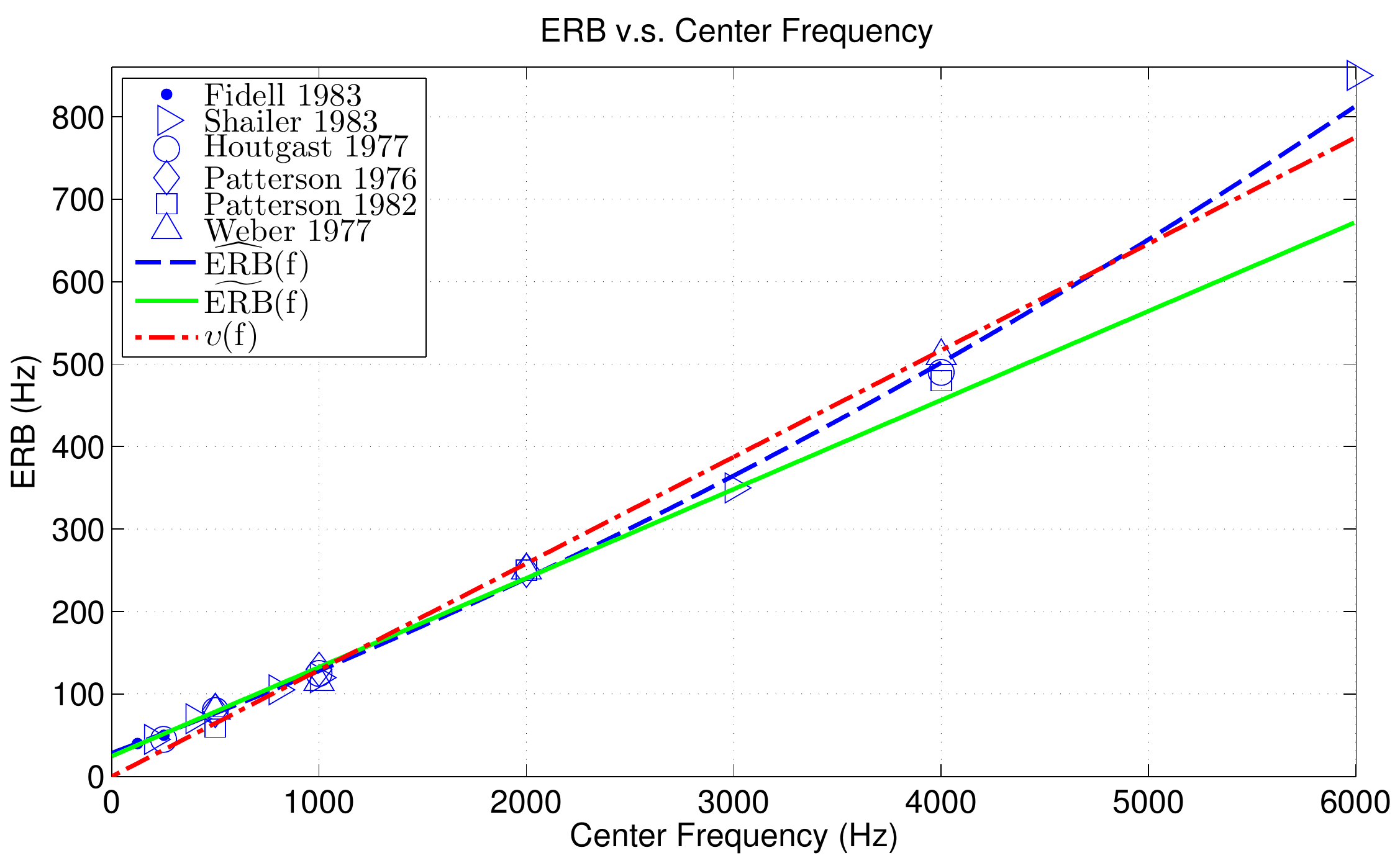}
\caption{Measured equivalent rectangular bandwidth versus center frequency, and ERB curves. }
\label{ERBvsCenterFrequency}
\end{figure}

From (\ref{eq:ERBSfunc}) and (\ref{eq:logfunc}) we have a new approximation of the ERB:
\begin{equation} \label{eq:scaling}
\begin{aligned}
\upsilon(f) = & 1 \big/ {\frac{d \Upsilon(f)}{df}} \\
= & \frac{\ln B}{A} \cdot f . 
\end{aligned}
\end{equation}
Choosing natural logarithm, i.e. $B =  e$, where $e = 2.718...$, we can get $A $ from linear fitting of experimental readings from the literature \cite{patterson1976auditory, weber1977growth, houtgast1977auditory, patterson1982deterioration, fidell1983effective, shailer1983gap} as shown in Fig.~\ref{ERBvsCenterFrequency}. 
We can see that 
\begin{equation} \label{eq:upsilon}
\upsilon(f) = \frac{1}{A}\cdot f ,
\end{equation} 
where $A = 7.7$ fits the data well. Then we have 
\begin{equation} \label{eq:Upsilon}
\Upsilon(f) = 
\begin{cases}
A \ln (f) + C , ~ f > f_m \\
0, ~ 0 \leq f \leq f_m 
\end{cases}  , 
\end{equation}
where $C = -23.1$. 

Equations (\ref{eq:upsilon}) and (\ref{eq:Upsilon}) are the proposed new ERB and ERBS functions.
Note here that the ERB of human auditory system may vary with age and sound level and from one listener to another \cite{moore1983suggested}. Thus the precise values of $A$ and $C$ may vary. 
However, the derivation from (\ref{eq:logfunc}) to (\ref{eq:etaCb}) shows that, as long as the ERB function has the proposed form of (\ref{eq:scaling}) or (\ref{eq:upsilon}), the resulting frequency scaling always satisfies the frequency coverage as (\ref{eq:etaCb}) shows. 

\begin{figure}[!h]
\centering
\includegraphics[width=0.5\textwidth]{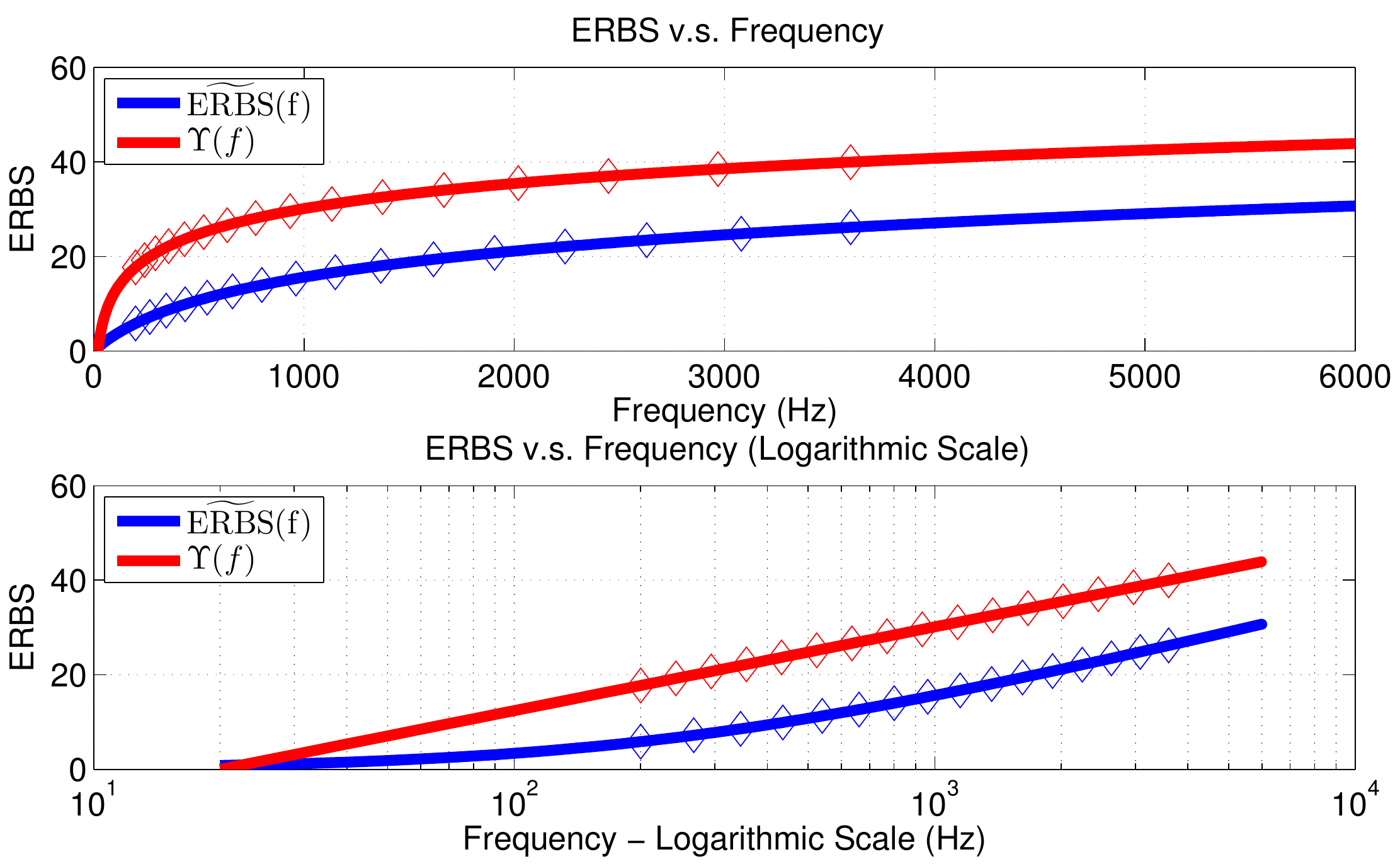}
\caption{The existing and proposed ERBS and corresponding selected center frequencies. }
\label{ERBSvsCenterFrequency}
\end{figure}

The existing and proposed ERBS functions are plotted in Fig.~\ref{ERBSvsCenterFrequency}. 
We can see that the proposed scaling follows the proposed logarithmic scaling, and is steeper at frequencies lower than about $1000$Hz. 
In this section we use $f_{min} = 200$Hz and $f_{max}=3600$Hz. The center frequencies that correspond to equidistant points on respective ERBS for $N_b=16$ are plotted in $\diamond$. We can see that  the proposed ERBS has more points at low frequencies. 
This can provide better frequency resolution on the lower frequencies as most of speaker fundamental frequencies are below $500$Hz, and usually most speech energies are in the fundamental frequency or its lower order harmonics \cite{deller1993discrete}.

\subsection{Frequency Coverage of the Gammatone Filterbank}

The frequency coverage $\eta_C^{(b)}$ is the property that we propose for the selection of center frequencies of an auditory filterbank. Here we use the gammatone filter to demonstrate the feature.

We can see from \cite{holdsworth1988implementing} that bandwidth of the gammatone filter is only dependent on the filter order $n$ ($n \geq 1$) and the ERB, i.e.
\begin{equation} \label{eq:fBgammatone}
f_{B,\Gamma}^{(b)}  
= k(n) \cdot \widetilde{\mathrm{ERB}}(f_C^{(b)}) ,
\end{equation} 
where
\begin{equation}
k(n) = 2 \sqrt{2^{1/n}-1} \cdot 
\Big[ \frac{\pi (2n-2)!2^{-(2n-2)}}{(n-1)!^2} \Big]^{-1} .
\end{equation}
This satisfies the assumptions in (\ref{eq:etaCb}) and (\ref{eq:constantK}).
Thus using the new ERB function (\ref{eq:upsilon}) instead of (\ref{eq:ERB}) in (\ref{eq:fBgammatone}), we have the Q-factor for the gammatone filter
\begin{equation} \label{eq:etaGamma}
\eta_{B,\Gamma}^{(b)} = \frac{A}{ k(n) } ,
\end{equation} 
which is constant over frequency, e.g. when $n=4$, we have $k(4)=0.8865$, and $\eta_{B,\Gamma}^{(b)} = 8.69 $. Thus given $N_b=16$, we can get $\eta_C^{(b)} \equiv 0.6$ from (\ref{eq:etaCb}), and $\eta_{C,\hat{\upsilon}}^{(b)} \equiv 0.66$ from (\ref{eq:etaCpsi}).

\begin{figure}[!h]
\centering
\includegraphics[width=0.5\textwidth]{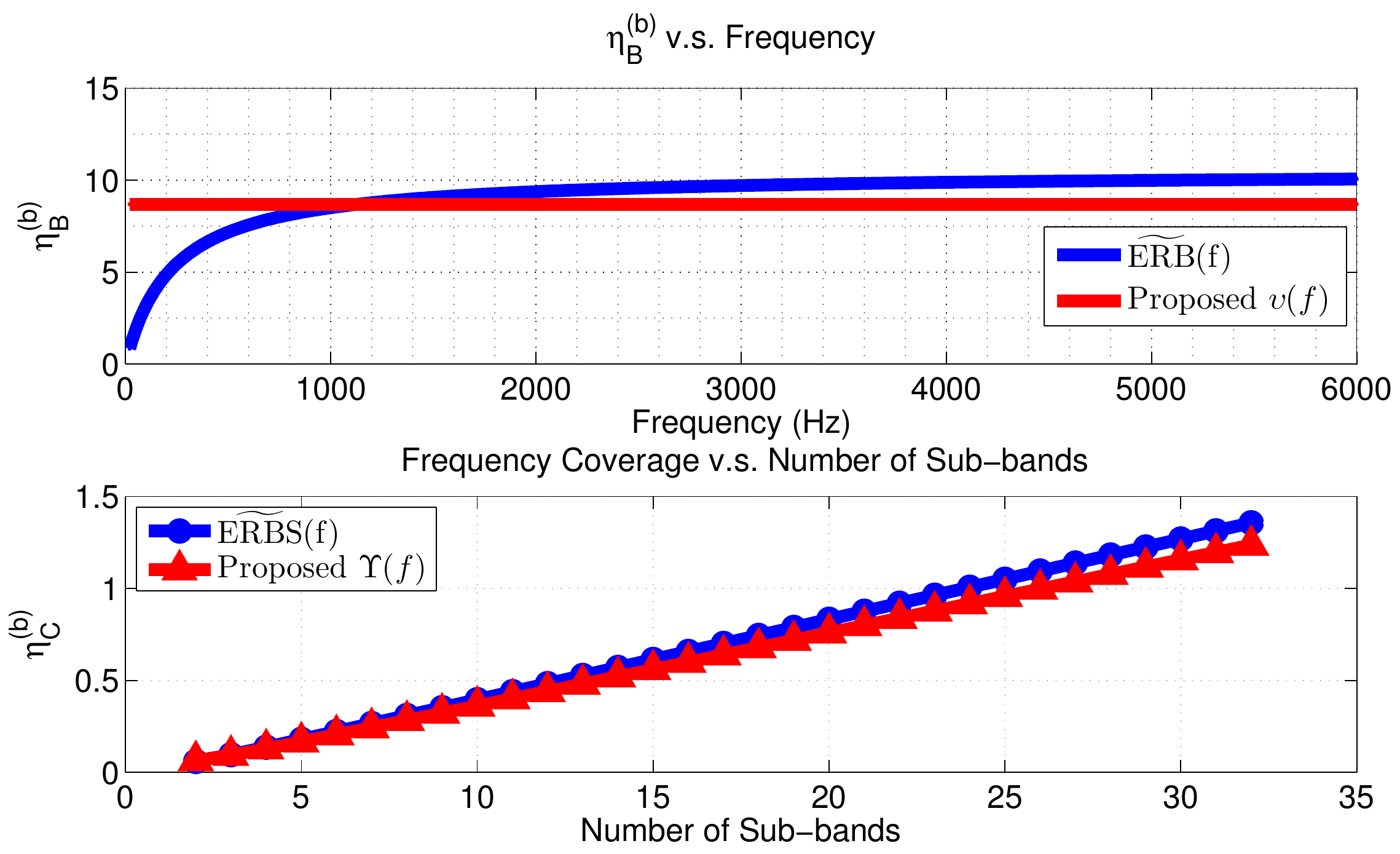}
\caption{Q-factor and frequency coverage for a 4-th order gammatone filterbank. }
\label{FreqCov_plots}
\end{figure}
Fig.~\ref{FreqCov_plots} further provides the frequency coverage of the proposed and existing ERBS over the number of subbands of the $4$-th order gammatone auditory filterbank for the frequency range of $[200, 3600]$Hz.
We can see from the top panel that for frequencies above about $500$Hz, both ERBs align well with each other. However, the existing ERB has decreasing Q-factors as frequencies decrease below about $500$Hz, while the proposed ERB is consistent across the entire frequency range. 
We can also see from the bottom panel that for both ERB scaling functions, the frequency coverage is constant for a given number of subbands $N_b$, and increases almost linearly with the number of subbands. The frequency coverage reaches about 1 at $N_b=24$ for both scaling. 
However, it can also be noted that for the same frequency range, the ERBS requires less number of subbands than the new logarithmic scale, for  a desired frequency coverage.




\section{Conclusions}

This paper investigates the frequency scaling of the auditory filterbanks, and proposes a novel frequency coverage metric for the selection of center frequencies of auditory filterbanks. 
We also propose a new ERB that aligns with the logarithmic frequency scaling, and derive that equidistant frequencies on the logarithmic frequency scale provide a consistent frequency coverage for the filterbanks. 
Moreover, we show that the existing and any possible linear ERB can also provide consistent frequency coverage.
The suggested frequency coverage is demonstrated using the gammatone filterbank.

%
%
\section*{Acknowledgment}
The author would like to acknowledge the contribution of the Australian Postgraduate Award and Australian Government Research Training Program Scholarship in supporting this research.
Due thanks are given to Professor S. Nordholm and anonymous reviewers for the review comments on early revisions of the manuscript. 
%

\bibliographystyle{IEEEtran}
\bibliography{Bibliography}

\end{document}